\documentclass[seceq]{jpsj3}

\usepackage{txfonts}
\usepackage{amsmath,amssymb}
\usepackage{graphicx}
\usepackage{subfigure}
\usepackage{bm}
\usepackage{color}

\newcommand{\nn}{\nonumber}

\newcommand{\bmm}{{\bm m}}
\newcommand{\LD}{\mathrm{LD}}

\title{$1/f$ Spectrum and 1-Stable Law in One-Dimensional Intermittent Map with Uniform Invariant Measure and Nekhoroshev Stability}

\author{Soya Shinkai\thanks{Present address: Department of Mathematical and Life Science, Graduate School of Science, Hiroshima University, Kagamiyama, Higashi-Hiroshima 739-8526, Japan.} and Yoji Aizawa}
\inst{\address{Department of Applied Physics, Faculty of Science and Engineering, Waseda University, 3-4-1 Okubo, Shinjuku-ku, Tokyo 169-8555}
}
\abst{
We investigate ergodic properties of a one-dimensional intermittent map that has not only an indifferent fixed point but also a singular structure such that a uniform measure is invariant under mapping.
The most striking aspect of our model is that stagnant motion around the indifferent fixed point is induced by the log-Weibull law, which is derived from Nekhoroshev stability in the context of nearly-integrable Hamiltonian systems.
Using renewal analysis, we derive a logarithmic inverse power decay of the correlation function and a $1/\omega$-like power spectral density.
We also derive the so-called 1-stable law as a component of the time-average distribution of a simple observable function.
This distributional law enables us to calculate a logarithmic inverse power law of large deviations.
Numerical results confirm these analytical results.
Finally, we discuss the relationship between the parameters of our model and the degrees of freedom in nearly-integrable Hamiltonian systems.
}

\kword{
$1/f$ spectrum, intermittent maps, Nekhoroshev stability, log-Weibull law, 1-stable law, renewal analysis
}

\begin{document}
\maketitle

\section{Introduction}
One of the most complex phenomena in chaotic dynamical systems is slow dynamics such as the $1/f$ spectral fluctuations.
Typical examples of such $1/f$ spectrum are often observed not only in dissipative systems through the saddle-node bifurcation~\cite{Manneville80} but also in conservative systems such as chaotic Hamiltonian systems~\cite{Kohyama84, GeiselZacherlRadons87}.
An aspect of these orbits in slow dynamics is an intermittent behavior.
Recently, a theory of intermittency has been developed toward non-hyperbolic dynamical systems using infinite ergodic theory~\cite{Aaronson97}, in which the indifferent fixed points play an essential role inducing the measure-theoretical singularity;
the polynomial decay of the correlation function, the $1/f$ spectrum, and the limit theorem for the time average of certain observable functions have been elucidated~\cite{Aizawa84, ShinkaiAizawa06, AkimotoAizawa07,Akimoto08}.
On the other hand, in spite of many reports about slow dynamics in chaotic Hamiltonian systems~\cite{Karney83ChirikovShepelyansky84MackayMeissPercival84MeissOtt85UmbergerFarmer85}, ergodic properties as well as the geometric structures have not yet been completely defined except in the cases of mushroom billiards~\cite{Miyaguchi07} and the two-dimensional piecewise linear map~\cite{AkaishiShudo09}.
One of the successes of the measure-theoretical approach for {\em Nekhoroshev stability}~\cite{footnoteNekhoroshev,Nekhoroshev77} is the {\em log-Weibull law}~\cite{Aizawa89a,Aizawa89b}:
for a sufficiently small perturbation parameter $\varepsilon$,
if the Nekhoroshev stability time is regared as a trapping time around a torus and the natural measure around tori is taken to be the Lebesuge measure,
the probability density of the first escape time $m$ around a torus is given by
\begin{equation}\label{log_Weibull}
	f_{\mathrm e}(m) = O \left( m^{-1} (\log m)^{-(1+d(s))} \right) \quad \mbox{as} \quad m \to \infty,
\end{equation}
where $s$ stands for degrees of freedom, and $d(s) = c_1 (s-1) /c_2$.
A similar estimate was discussed in a study of a perturbation theory for nearly-integrable Hamiltonian systems~\cite{MorbidelliVergassola97}.
The induction phenomena in the lattice vibration~\cite{Aizawa89b} and the universal long time tail of clustering motions in $N$-body systems~\cite{Aizawa00b,AizawaSatoIto00} agree well with the log-Weibull law.

In studies of intermittency, the Pikovsky map~\cite{Pikovsky91} was the first model equipped with slow dynamics and a uniform invariant measure, which corresponds to the property that the Liouville measure is invariant under a Hamiltonian flow.
The map's thermodynamic formalism and anomalous transport have been studied using periodic orbit theory~\cite{ArtusoCristadoro04}.
However, since the map is defined implicitly, it is inadequate for discussing dynamical properties.
Miyaguchi and Aizawa~\cite{MiyaguchiAizawa07} improved the Pikovsky map via a piecewise-linearization and defined it explicitly.
They rigorously investigated its spectral properties using the Frobenius-Perron operator.
In 2010, Cristadoro {\em et al.}~\cite{Cristadoro10} derived ergodic properties such as the polynomial decay of the correlation function and stable laws using statistical properties arising from the Pikovsky map.
However, for both maps, extremely strong intermittency around fixed points such as Nekhoroshev stability have not yet been considered.
Therefore, our first aim is to construct a one-dimensional intermittent map with a uniform invariant measure and Nekhoroshev stability.

For hyperbolic dynamical systems, both an exponential decay of large deviations and the rate function exactly determine the thermodynamic formalism or the statistical mechanics described by the Gibbs measure~\cite{Ellis85}.
However, many problems exist with the statistical mechanics of the weak Gibbs measure in intermittent dynamical systems~\cite{Yuri03}.
In 2009, Melbourne~\cite{Melbourne09}, Pollicott and Sharpe~\cite{PollicottSharp09} independently demonstrated polynomial decays of large deviations for a certain class of observable functions in the Pomeau-Manneville map.
Artuso and Manchein~\cite{ArtusoManchein09} also numerically showed polynomial decays in slowly mixing dynamical systems.
One of our objectives is to obtain estimates of large deviations in our model.

This paper is organized as follows.
In \S\ref{s2}, we introduce a class of one-dimensional intermittent maps in which a uniform measure is approximately invariant.
Next we show that the map has Nekhoroshev stability around the indifferent fixed point.
In addition, using renewal analysis, we derive the correlation function and power spectral density.
In \S\ref{s3}, we present a limit theorem of the time average for a simple observable function and derive the decay rate of large deviations from the limit theorem.
In \S\ref{s4}, we present our numerical results.
Finally, \S\ref{s5} is devoted to a summary and discussion.

\section{\label{s2}The Model and its Statistical Features}
\subsection{Map equipped with a uniform invariant measure}\label{ss2.1}
Here we consider the map $T$ defined in the interval $[0,1]$ as
\begin{equation}\label{def_map_T}
	T(x) = \left\{
	\begin{array}{lcl}
		T_0(x) = x + (1-a) \, g\left( \frac{x}{a} \right) & \mbox{for} & x \in [0,a),\\
		T_1(x) = x - a + a \, g^{-1}\left( \frac{x-a}{1-a} \right) & \mbox{for} & x \in [a,1],
	\end{array}
	\right.
\end{equation}
where $a$ is a constant ($0<a<1$) and function $g: [0,1] \to [0,1]$ has the following properties:
\begin{itemize}
\item
	the inverse function $g^{-1}$ exists;
\item
	$g(0) = 0$ and $g(1) = 1$;
\item
	for $t \ll 1$, $g(t) \ll t$ $\left( g^{-1}(t) \gg t \right)$ and $g'(t) \ll 1$ $\left( \left( g^{-1} \right)'(t) \gg 1 \right)$.
\end{itemize}

Given these properties for $g$, the uniform density can be approximately derived as a solution of the Frobenius-Perron equation of the map $T$ as follows.
Let $b \, (< a)$ be some small constant.
For the points $y_0 \in [0,a)$ and $y_1 \in [a,1]$ that satisfy $x = T(y_j) \, (j=0,1)$,
if $y_0 /a \ll 1$ and $(y_1-a)/(1-a) \ll 1$ for small $x < b$,
we can approximately write the Frobenius-Perron equation as
\begin{eqnarray}
	\rho(x) & = & \sum_{y_j \in T^{-1}x} \frac{\rho(y_j)}{T'(y_j)} \nn\\
	 & = & \rho(y_0) \left\{ 1 + \frac{1-a}{a} g'\left( \frac{y_0}{a} \right) \right\}^{-1}
	 + \rho(y_1) \left\{ 1 + \frac{a}{1-a} \left(g^{-1}\right)' \left( \frac{y_1-a}{1-a} \right) \right\}^{-1} \nn\\
	 & \approx & \rho(y_0) \left\{ 1 - \frac{1-a}{a} g'\left( \frac{y_0}{a} \right) \right\}
	 + \rho(y_1) \left\{ \frac{a}{1-a} \left(g^{-1}\right)' \left( \frac{y_1-a}{1-a} \right) \right\}^{-1}, \label{approFP}
\end{eqnarray}
where $\rho$ is the invariant density of the map $T$.
We can also obtain the approximate relations
\begin{equation*}
	x \approx y_0 \quad \mbox{and} \quad g(x/a) \approx \frac{y_1-a}{1-a}.
\end{equation*}
Substituting these expressions and the formula for differentiation of an inverse function
\begin{equation*}
	\left( g^{-1} \right)' \left( g(t) \right) = \left\{ g'(t) \right\}^{-1}
\end{equation*}
into eq.~(\ref{approFP}), we obtain for small $x < b$
\begin{equation}\label{FPsmallx}
	\rho(x) \approx \rho(y_0) \left\{ 1 - \frac{1-a}{a} g' \left( \frac{x}{a} \right) \right\}
	+ \rho(y_1) \frac{1-a}{a} g' \left( \frac{x}{a} \right).
\end{equation}
For $x>b$, we assume that the slopes of $T_0$ and $T_1$ can be approximately equal to $(1-b) \big/ (a-b)$ and $(1-b) \big/ (1-a)$, respectively.
Then, the Frobenius-Perron equation can be approximated by
\begin{equation}\label{FPfinitex}
	\rho(x) \approx \rho(y_0) \, \frac{a-b}{1-b} + \rho(y_1) \, \frac{1-a}{1-b}.
\end{equation}
Therefore, eqs.~(\ref{FPsmallx}) and (\ref{FPfinitex}) imply that the uniform density $\rho(x)=1$ is an approximate solution of the Frobenius-Perron equation of $T$ for $x \in (0,1)$.

\begin{figure}
	\centering
	\subfigure[]{
		\includegraphics[width=0.45\hsize]{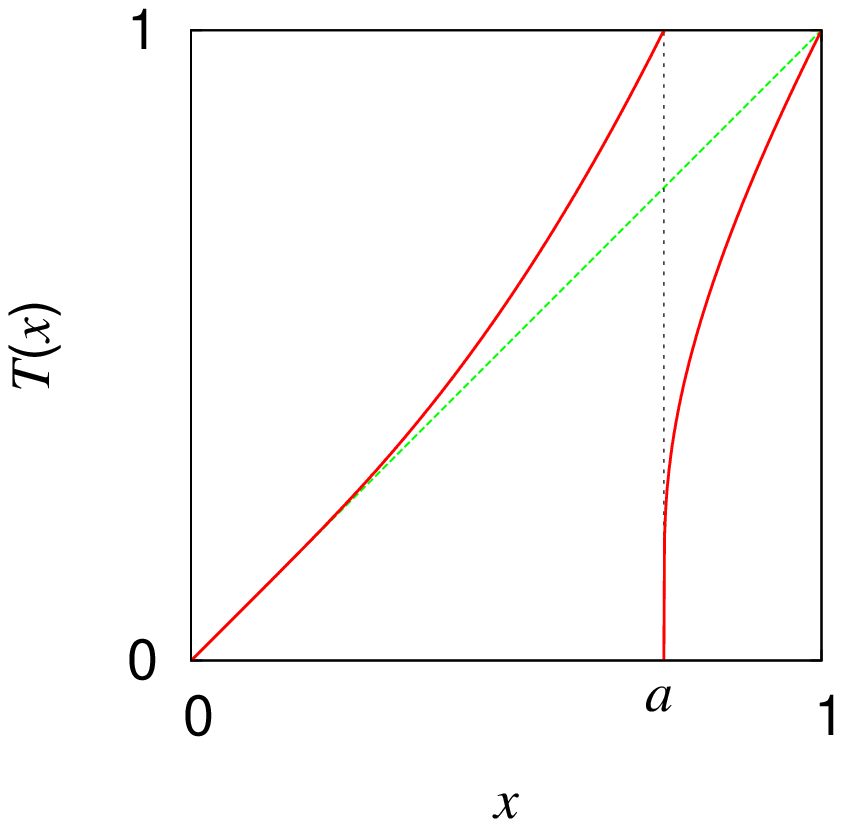}
		\label{fig1a}
	}
	\quad
	\subfigure[]{
		\includegraphics[width=0.45\hsize]{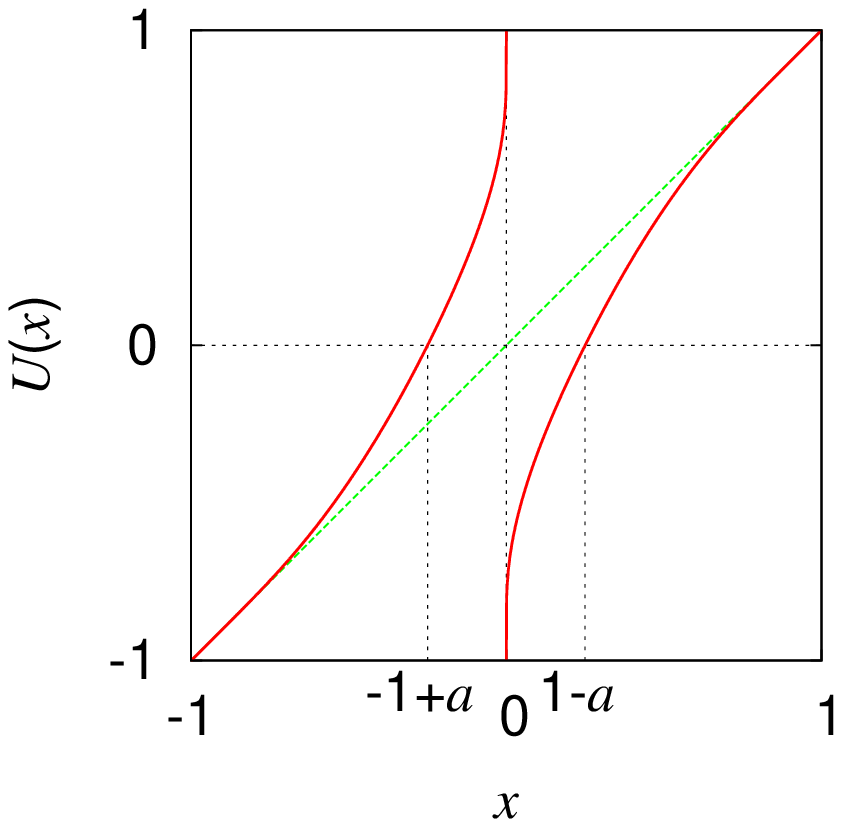}
		\label{fig1b}
	}
	\caption{(Color online) For $\beta = 1$, (a) a graph of the map $T(x)$  and (b) a graph of the map $U(x)$.}
\end{figure}

In what follows, we consider the function
\begin{equation}\label{def_g}
	g(t) = t^{1+\beta} \, \exp \left( 1 - t^{-\beta} \right),
\end{equation}
where the parameter $\beta > 0$.
The function obviously satisfies the three conditions above~\cite{footnote1}.
In the study of infinite ergodic theory, this function type was introduced by Thaler~\cite{Thaler83},
and its extremely intermittent characteristics were discussed by Shinkai and Aizawa~\cite{ShinkaiAizawa08}.
In the context of deterministic diffusion, such a function type generates logarithmic growth of the mean squared displacement~\cite{DragerKlafter00}.
As shown in Appendix\ref{AppA}, we can derive its inverse function explicitly,
\begin{equation}\label{def_g_inverse}
	g^{-1}(t) = \left\{ \eta \, W\left( \eta^{-1} (e/t)^{1/\eta} \right) \right\}^{-1/\beta},
\end{equation}
where
\begin{equation}\label{def_eta}
	\eta = 1 + \frac{1}{\beta},
\end{equation}
and the function $W$ is called the {\em Lambert} $W$ {\em function}~\cite{Corless96} and is defined as the solution of the equation $z = W(z) \, e^{W(z)}$.
The parameter $a$ is defined so that the derivative $T'(x)$ is the same at $x = a$ and $x = 1$:
\begin{equation*}
	a \equiv \frac{2\beta + 1}{2\beta+2}.
\end{equation*}
The map $T$ for $\beta =1$ is shown in Fig.~\ref{fig1a}.
The left side, $T_0(x)$, has a structure similar to  that of Pomeau-Manneville-type intermittent maps;
the origin is an indifferent fixed point satisfying $T_0(0)=0$ and $T'_0(0) = 1$.
However, the derivative of the right side, $T'_1(x)$, is divergent at $x=a$.

In calculating the correlation function and power spectral density, we use the symmetrized map $U$ defined on the interval $[-1,1]$ as follows:
\begin{equation*}
	U(x) = \left\{
	\begin{array}{lclcl}
		T_0(x+1) -1 & = &
		x + (1-a) \, g\left( \frac{x+1}{a} \right) & \mbox{for} & x \in [-1,-1+a),\\
		-T_1(-x+a) + 1 & = &
		x + 1 - a \, g^{-1}\left( \frac{-x}{1-a} \right) & \mbox{for} & x \in [-1+a,0),\\
		T_1(x+a) - 1 & = &
		x - 1 + a \, g^{-1}\left( \frac{x}{1-a} \right) & \mbox{for} & x \in [0,1-a),\\
		-T_0(-x+1) + 1 & = &
		x - (1-a) \, g\left( \frac{-x+1}{a} \right) & \mbox{for} & x \in [1-a,1].
	\end{array}
	\right.
\end{equation*}
Figure~\ref{fig1b} shows a graph of the map $U$ for $\beta=1$.

\begin{figure}
	\centering
	\includegraphics[width=\hsize]{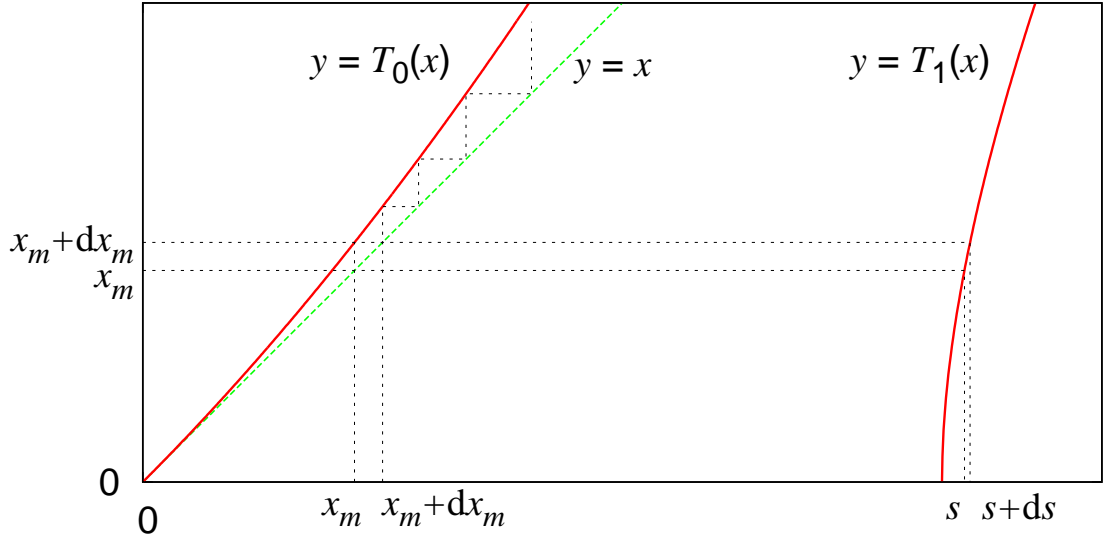}
	\caption{(Color online) Points $x_m$ and $s$ satisfy $T_0^m(x_m) = 1$ and $T_1(s)=x_m$.
	The image $T_1\left( [s,s+ds) \right)$ corresponds to the interval $[x_m,x_m+dx_m)$.}
	\label{fig2}
\end{figure}

\subsection{Distributions of first escape and residence times}\label{ss2.2}
Here we consider the {\em first escape time} from the interval $[0,a)$ and the {\em residence time} in the interval and then asymptotically estimate each probability density function.

Let us consider the points $x_m \in [0,a)$ and $s \in [a,1]$ such that $T_0^m(x_m) = 1$ and $T_1(s) = x_m$ (see Fig.~\ref{fig2}).
First, we estimate the relationship between $x_m$ and $m$ as follows:
For $\Delta x_m \equiv T_0(x_m) - x_m$ and $\Delta m \equiv m - (m-1)$,
eq.~(\ref{def_map_T}) can be written as
\begin{equation}\label{differenceEq}
	\frac{\Delta x_m}{\Delta m} = (1-a) \, g(x_m/a).
\end{equation}
If we use the continuous approximation for $x_m \ll 1 \, (m \gg 1)$, eq.~(\ref{differenceEq}) becomes
\begin{equation}\label{dx/dm}
	\frac{d x}{d m} \approx (1-a) \, g(x/a).
\end{equation}
Now we integrate eq.~(\ref{dx/dm}) as follows:
\begin{eqnarray}\label{mG}
	m = \int_0^m \, d m' & \approx & \frac{1}{1-a} \int_{x_m}^a \frac{d x}{g(x/a)}\nn\\
	& = & \frac{a}{1-a} \int_{x_m/a}^1 \frac{d y}{g(y)} \qquad (y=x/a)\nn\\
	& = & \frac{a}{1-a} \left\{ G(x_m/a) - G(1) \right\},
\end{eqnarray}
where the function $G(x)$ is the indefinite integral $-\int^x d y / g(y)$.
Let us assume that $G(x_m/a) \gg G(1)$ for $x_m/a \ll 1$, and that the function $G$ has an inverse $G^{-1}$.
Then, eq.~(\ref{mG}) can become $m = O\left( G(x_m/a) \right)$ as $x_m \to 0 \, (m \to \infty)$, and we can obtain the relationship
\begin{equation}\label{x_m-m}
	x_m / a = O\left( G^{-1}(m) \right) \quad \mbox{as} \quad m \to \infty.
\end{equation}
For initial points uniformly distributed on the interval $[0,a)$,
an orbit starting in the interval $[x_m, x_m+dx_m)$ spends $m$ time-steps to escape from $[0,a)$.
Therefore
the first escape time probability density $f_{\mathrm e}$ is defined as
\begin{equation}\label{escape_measure}
	f_{\mathrm e} (m) \, dm \equiv d x_m.
\end{equation}
Using eqs.~(\ref{dx/dm}), (\ref{x_m-m}), and (\ref{escape_measure}), we can derive the formula
\begin{equation}\label{estimation_es_dis}
	f_{\mathrm e}(m)  =  \frac{dx_m}{dm} = O \left( g \left( G^{-1}(m) \right) \right) \quad \mbox{as} \quad m \to \infty.
\end{equation}

Next, we consider the probability of orbits being injected into the interval $[x_m,x_m+d x_m)$.
Under the continuous approximation,
we assume that the interval $[s,s+d s)$ is mapped onto the interval $[x_m,x_m+d x_m)$, as shown in Fig.~\ref{fig2}.
An orbit that starts in the interval $[x_m,x_m+d x_m)$ resides for $m$ time-steps in the interval $[0,a).$
Therefore, using the invariant measure $\rho$, the residence time probability density $f_{\mathrm r}$ is defined as
\begin{equation}\label{res_measure}
	f_{\mathrm r}(m) \, dm \equiv \rho(s) \, ds.
\end{equation}
From the definition of point $s$ and under the condition that $g^{-1}\left( \frac{s-a}{1-a} \right) \gg \frac{s-a}{1-a}$,
we can estimate
\begin{equation*}
	g(x_m/a) \approx \frac{s-a}{1-a}.
\end{equation*}
Using eqs.~(\ref{dx/dm}), (\ref{x_m-m}), (\ref{res_measure}), and the approximation $\rho(s) \approx 1$,
we can estimate the residence time probability density as
\begin{eqnarray}\label{estimation_res_dis}
	f_{\mathrm r}(m) & = & O \left( g'(x_m/a) \cdot \frac{d x_m}{d m} \right) \nn\\
	& = & O \left(g'\left(G^{-1}(m)\right) \cdot g\left( G^{-1}(m) \right) \right) \quad \mbox{as} \quad m \to \infty.
\end{eqnarray}
Note that the relationship between this and the first escape time probability density is $f_{\mathrm r}(m) = O\left( g'\left(G^{-1}(m)\right) \right) \cdot f_{\mathrm e}(m)$.
In other words, the coefficient $g'\left(G^{-1}(m)\right)$ is the result of a uniform invariant measure and the orbits injected into near the origin.

If we use formulas (\ref{estimation_es_dis}) and (\ref{estimation_res_dis}) for $g(t)$ as defined by eq.~(\ref{def_g}), the probability density functions become
\begin{eqnarray}
	f_{\mathrm e} (m) & = & O \left( m^{-1} \, (\ln m)^{-\eta} \right) \qquad (m \gg 1), \label{pdf_first_escape}\\
	f_{\mathrm r} (m) & = & O \left( \left\{ 1 + \eta \, (\ln m)^{-1} \right\} \, m^{-2} \, (\ln m)^{-\eta}
	\right) \qquad (m \gg 1). \label{pdf_residence}
\end{eqnarray}
The derivation of these asymptotic relations is given in Appendix\ref{AppB}.
As mentioned in the Introduction, $f_{\mathrm e}(m)$ obeys the log-Weibull law.
In other words, the function $g$ defined by eq.~(\ref{def_g}) induces Nekhoroshev stability around the origin.

\subsection{Correlation function and power spectral density}\label{ss2.3}
Renewal theory~\cite{Cox62,Feller71} enables us to calculate the correlation function of coarse-grained orbits under the symmetric map $U$,
\begin{equation*}
	C(n) \equiv \left\langle \sigma(x) \, \sigma(U^n x) \right\rangle_\rho,
\end{equation*}
where the function $\sigma$ is defined by
\begin{equation*}
	\sigma(x) = \left\{
	\begin{array}{ccl}
		- 1 & \mbox{ for } & x < 0,\\
		1   & \mbox{ for } & x > 0,
	\end{array}
	\right.
\end{equation*}
and $\langle \cdot \rangle_\rho$ stands for the ensemble average under the uniform invariant measure.
For an orbit $\{ x_n \} \equiv \{ U^n x \}$, the $j$-th event occurs at the renewal time $n_j$ ($\sigma(x_{n_{j}-1}) \cdot \sigma(x_{n_j}) = -1$).
As shown in the previous subsection, the intervals of time between successive renewal events, $m_2 = n_2-n_1$, $m_3 = n_3 - n_2$, $\cdots$, can be independent and identically distributed (i.i.d.) random variables with density $f_{\mathrm r}(m)$;
the first interval of time $m_1 = n_1$ can be a random variable with density $f_{\mathrm e}(m)$.
Thus, the relationship between $f_{\mathrm e}(m)$ and $f_{\mathrm r}(m)$ satisfies the special form
\begin{equation}\label{relationship_fe_fr}
	f_{\mathrm e}(m) = O \left( \int_m^\infty f_{\mathrm r}(\tau) \, d\tau \right) \quad \mbox{as} \quad m \to \infty.
\end{equation}
This implies that the renewal process under the initial uniform ensemble is an {\em equilibrium renewal process}~\cite{Cox62}.

The Laplace transform $C^\ast(z) = \int_0^\infty e^{-zn} C(n) \, dn$ satisfies
\begin{equation}\label{Laplace_correlation}
	C^\ast(z) = \frac{1}{z} - \frac{2}{\langle m \rangle_{\mathrm r} z^2} \,
	\frac{1-f_{\mathrm r}^\ast(z)}{1+f_{\mathrm r}^\ast(z)},
\end{equation}
where $\langle m \rangle_{\mathrm r} = \int_0^\infty m f_{\mathrm r}(m) \, dm$, and $f_{\mathrm r}^\ast(z)$ is the Laplace transform of $f_{\mathrm r}(m)$~\cite{Aizawa84,AkimotoAizawa07}.
Using the Tauberian theorem~\cite{Feller71}, which states that the asymptotic behavior of a function at infinity uniquely determines the behavior of its Laplace transform near the origin and vice versa, we can estimate the correlation function $C(n)$ as $n \to \infty$.
Moreover, eq.~(\ref{Laplace_correlation}) enables us to estimate the power spectral density $S(\omega)$ of the orbits $\left\{ \sigma(x_n) \right\}$, which is usually defined by the real part of $C^\ast(i\omega)$ according to the Wiener-Khinchin theorem.

As shown in Appendix\ref{AppB}, we can summarize as follows:
\begin{eqnarray}
	C(n) & = & O \left( (\ln n)^{-1/\beta} \right) \quad  \mbox{as} \quad n \to \infty, \label{correlation}\\
	S(\omega) & = & O \left( \omega^{-1} \, \left\{ r(\omega) \right\}^{-1/\beta} \,
	\sin\left( \theta(\omega) / \beta \right) \right) \quad \mbox{as} \quad \omega \to 0 , \label{psd}
\end{eqnarray}
where the functions are defined by
\begin{equation}\label{r_theta_omega}
	r(\omega) = \sqrt{(\ln \omega)^2 + (\pi/2)^2} \quad \mbox{and} \quad
	\theta(\omega) = \tan^{-1} \left( -\frac{\pi}{2 \ln \omega}  \right).	
\end{equation}
Equations (\ref{psd}) and (\ref{r_theta_omega}) imply that the power spectral density has the special form $1 / \omega$ with the logarithmic correction term,
\begin{equation}\label{eq:psd}
	S(\omega) = O \left( \omega^{-1} (- \ln \omega)^{-1/\beta} \right) \quad \mbox{as} \quad \omega \to 0 .
\end{equation}

\section{Distribution of Partial Sums}\label{s3}
According to ergodic theory, the set $A=[0,1]$ satisfies $T^{-1} A = A$, and its measure is trivially $\mu(A)=1$ for the {\em finite} invariant measure $\mu$.
Hence, the map $T$ is ergodic.
Then, Birkhoff's pointwise ergodic theorem ensures the limit law
\begin{equation*}
	\lim_{n\to\infty} \frac{S_n(\phi)}{n} = \int_0^1 \phi \, d\mu \quad \mbox{for} \quad \mbox{a.e.} \quad x \in A, \quad \forall \phi \in L^1(\mu),
\end{equation*}
where $S_n(\phi) \equiv \sum_{k=0}^{n-1} \phi \left( T^k x \right)$ stands for the partial sum of the observable function $\phi$.
However, problems still remain regarding physical measurements for sufficiently large $n$:
What distribution does the time average $S_n(\phi) \big/ n$ obey?
How does it converge to the phase average?

For {\em infinite} ergodic systems, Birkhoff's ergodic theorem is not applicable.
However, the so-called Darling-Kac-Aaronson theorem~\cite{Aaronson97}, which states that the rescaled partial sum for an $L^1_+$ observable function behaves as the Mittag-Leffler random variable, is a universal law.
The essence of its proof is to solve the asymptotic renewal equation for the Darling-Kac set.
The origin of such a renewal equation belongs to probability theory~\cite{Feller49DarlingKac57}.
By using renewal analysis for intermittent maps, it was elucidated that a family of stable distributions is a universal distributional law~\cite{Aizawa89cKikuchiAizawa90TanakaAizawa93}.

\subsection{Renewal analysis for distribution of partial sums}\label{ss3.1}
Here, for simplicity, we consider the observable function defined by
\begin{equation}\label{def_phi}
	\phi(x) = \left\{
	\begin{array}{lcl}
		0 & \mbox{ for } & x < a,\\
		1 & \mbox{ for } & x > a.
	\end{array}
	\right.
\end{equation}
This function enables us to calculate the classical renewal analysis.
Moreover, let us consider the set
\begin{equation*}
	A_n = [0, a_n],
\end{equation*}
where the point $a_n$ is defined by $T_0^{n}(a_n)=1$ for $n \ge 1$ and $a_0 = 1$.
From eqs.~(\ref{x_m-m}) and (\ref{G_Ginverse}), the measure of $A_n$ decays as follows:
\begin{equation}\label{muAn}
	\mu(A_n) = O\left( (\ln n)^{-1/\beta} \right) \quad \mbox{as} \quad n \to \infty.
\end{equation} 
As mentioned in \S\ref{ss2.3}, the intervals of time between successive renewal events $\bmm_1$ and $\{ \bmm_j \} \, (j \ge 2)$ can be i.i.d. random variables with density $f_{\mathrm e}$ and $f_{\mathrm r}$, respectively.
Note that there exists another successive time which is equal to the residence time in the interval $[a,1]$ and contributes to $\phi ( T^k x)= 1$.
In our definition, such a time is included in the intervals $\bmm_j \, (j \ge 2)$.
However, the time can be ignored in what follows, since the hyperbolic structure of the right side of the map $T_1(x)$ is expected to induce an exponential decay of correlations for the successive time.

Given the initial uniform ensemble on the interval $A=[0,1]$, for sufficiently large $n$, the ensemble is divided in two with respect to the behavior of $S_n$.
For an initial point $x \in A_n$, it is trivial that $S_n(\phi) = 0$ and $n < \bmm_1$.
Therefore, the probability density function of $S_n(\phi) \big/ n$ for $x \in A_n$ becomes
\begin{equation*}
	p_n(y) \equiv \Pr \left\{ y \le \frac{S_n(\phi)}{n} < y + dy \right\} \bigg/ dy = \mu(A_n) \, \delta(y).
\end{equation*}
On the other hand, for an initial point $x \in A \setminus A_n$, we can approximate
\begin{equation}\label{n_Sn_m}
	n - S_n(\phi) \approx \bmm_2 + \bmm_3 + \cdots + \bmm_{M+1}
\end{equation}
assuming that events occur $M+1$ times in $n$-time-steps.
Note that the number of times $M$ is a function of $n$ and behaves as a random variable.
However, for the equilibrium renewal process under the density $f_{\mathrm r}$, the Laplace transform becomes $M^\ast(z) = 1 \big/ \left( \langle m \rangle_{\mathrm r} \, z^2 \right)$ as $z \to 0$;
hence $\langle M(n) \rangle_\rho = n \big/ \langle m \rangle_{\mathrm r}$ as $n \to \infty$~\cite{Cox62}.
Then, the distribution of $S_n(\phi) \big/ n$ is given by
\begin{equation}\label{QtildeQ}
	Q_n(y) = \Pr \left\{ \frac{S_n(\phi)}{n} < y \right\} = 1 - \tilde{Q}_n(1-y),
\end{equation}
where $\tilde{Q}_n$ is the distribution of $\left( 1 - S_n(\phi) \big/ n \right)$, and the approximate relation~(\ref{n_Sn_m}) implies
\begin{equation*}
	\tilde{Q}_n(y) \approx \Pr \left\{ \sum_{j=1}^M \frac{\bmm_{j+1}}{n} < y \right\}.
\end{equation*}
Moreover, let $q_n(y)$ and $\tilde{q}_n(y)$ be the probability density functions of $Q_n$ and $\tilde{Q}_n$, respectively.
From eq.~(\ref{QtildeQ}), it is obvious that $q_n(y)$ is equal to $\tilde{q}_n(1-y)$.
In what follows, we will discuss $\tilde{q}_n$.
Using the result of renewal analysis $\langle M (n) \rangle_\rho = n / \langle m \rangle_{\mathrm r}$ for sufficiently large $n$ and the fact that $f_{\mathrm r}$ belongs to the domain of attraction of the 1-stable law~(see Appendix\ref{AppC}), the characteristic function of $\tilde{q}_n$ can be approximated as
\begin{equation}\label{qz}
	\tilde{q}^\ast_n(z) \equiv \int_{-\infty}^\infty e^{izy} \tilde{q}_n(y) \, dy \approx \left\{ \psi \left( \frac{z}{n} \right) \right\}^{n / \langle m \rangle_{\mathrm r}},
\end{equation}
where $\psi$ is the characteristic function of the 1-stable distribution.
Therefore, the probability density function of $S_n(\phi) \big/ n$ for $x \in A \setminus A_n$ becomes
\begin{equation*}
	p_n(y) = \left( 1 - \mu(A_n) \right) \, \tilde{q}_n(1-y).
\end{equation*}

Our renewal analysis for the distribution of the partial sum is summarized as follows:
For a uniformly distributed initial point $x \in A$,
\begin{equation}\label{dis_partial_sum}
	p_n(y) = \mu(A_n) \, \delta(y) + \left( 1 - \mu(A_n) \right) \, \tilde{q}_n(1-y),
\end{equation}
where $\mu(A_n)$ decays as eq.~(\ref{muAn}), and $\tilde{q}_n$ is the 1-stable probability density function with its characteristic function defined by eq.~(\ref{qz}).

\subsection{Large deviations}\label{ss3.2}
As mentioned in the Introduction, the behavior of large deviations is an important property for constructing the thermodynamic formalism.

Here we consider the following large deviations for initial points $x \in A \setminus A_n$:
\begin{equation*}
	\LD (n,\varepsilon) \equiv \Pr \left\{ \left| \frac{S_n(\phi)-n\langle \phi \rangle_\rho}{n}  \right| > \varepsilon \right\}.
\end{equation*}
From the approximate relation~(\ref{n_Sn_m}) and (\ref{qz}), we can derive
\begin{equation}\label{LD1}
	\LD (n,\varepsilon) \approx \Pr \left\{ \left| \frac{\sum_{j=1}^M \bmm_{j+1} - (1-\langle\phi\rangle_\rho) \langle m \rangle_{\mathrm r} M}{M} \right| >
	\langle m \rangle_{\mathrm r} \, \varepsilon \right\}.
\end{equation}
On the other hand, since the distribution of the random variables $\{ \bmm_{j+1} \}$ obeys the 1-stable distribution $V$ with the slowly varying function $(\ln x)^{-\eta}$ (see Appendix\ref{AppC}), Aaronson and Denker's theorem~\cite{AaronsonDenker98} states that there exists some constant $b_M$ such that
\begin{equation*}
	\Pr \left\{ \frac{\sum_{j=1}^M \bmm_{j+1} - b_M}{M \, (\ln M)^{-\eta}} \le y \right\} \to V(y) \qquad (M \to \infty).
\end{equation*}
If we apply the asymptotic property of 1-stable distributions~\cite{Feller71}, $1-V(y)+V(-y) \sim y^{-1}$, the following estimate can be derived for $y= \langle m \rangle_{\mathrm r} \, \varepsilon \, (\ln M)^\eta$:
\begin{equation}\label{LD2}
	\Pr \left\{ \left| \frac{\sum_{j=1}^M \bmm_{j+1} - b_M}{M} \right| >  \langle m \rangle_{\mathrm r} \, \varepsilon \right\}
	\sim \left\{ \langle m \rangle_{\mathrm r} \, \varepsilon \, \left( \ln M \right)^\eta \right\}^{-1}.
\end{equation}
By comparing eq.~(\ref{LD1}) with eq.~(\ref{LD2}), we can estimate the large deviations as
\begin{equation}\label{large_deviations}
	\LD (n,\varepsilon) = O \left( (\ln n)^{-\eta} \right)
\end{equation}
for sufficiently large $n$ and initial points $x \in A \setminus A_n$.

\section{Numerical Results}\label{s4}

\subsection{Entropy}\label{ss4.1}
\begin{figure}
	\centering
	\includegraphics[width=\hsize]{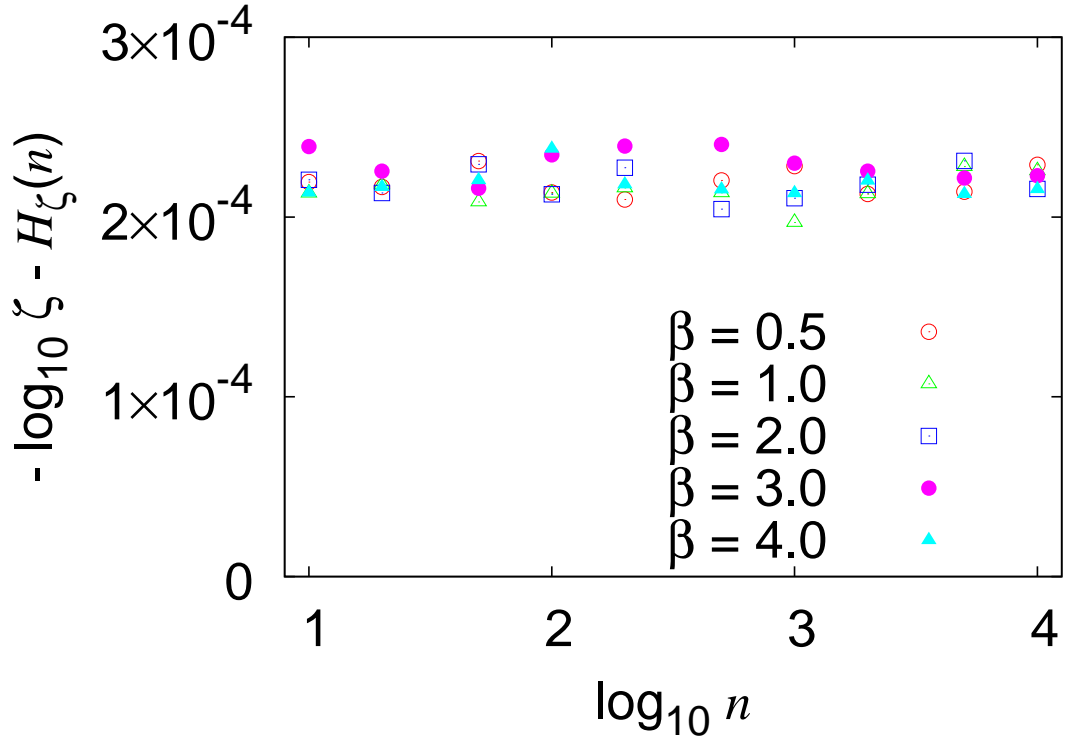}
	\caption{(Color online) Difference between $-\log_{10} \zeta$ and equipartition entropy $H_\zeta(n)$ for five different values of parameter $\beta$ and $\zeta = 10^{-3}$.
	For each $\beta$, $10^6$ initial points are uniformly distributed in the interval $(0,1)$.}
	\label{fig3}
\end{figure}

First, to check that the uniform measure is approximately invariant under mapping $T$, we calculate the equipartition entropy
\begin{equation*}
	H_\zeta(n) \equiv - \sum_{j = 1}^{\zeta^{-1}} \lambda_n(D_j) \log_{10} \lambda_n(D_j),
\end{equation*}
where $\zeta$ is the length of the set $D_j \equiv \left( (j-1) \zeta, j \zeta \right) \, (j = 1, 2, \cdots, \zeta^{-1})$,
$\lambda_n(D_j)$ denotes the probability measure of $D_j$ at time-step $n$,
and an initial ensemble is uniformly given on the interval $(0,1)$.
In ergodic theory, the upper bound of the entropy is $- \log_{10} \zeta$.
The numerical results are shown in Fig.~\ref{fig3} for $\zeta = 10^{-3}$ and five different values of parameter $\beta$, where $10^6$ initial points are uniformly distributed.
Because the relative error $(-\log_{10} \zeta - H_\zeta(n)) / (-\log_{10} \zeta)$ is less than $10^{-4}$, the uniform measure is approximately invariant under mapping $T$ in our simulations.

\subsection{Statistical features}\label{ss4.2}
\subsubsection{Distributions of first escape and residence times}
\begin{figure}
	\label{fig4}
	\centering
	\subfigure[]{
		\includegraphics[width=0.45\hsize]{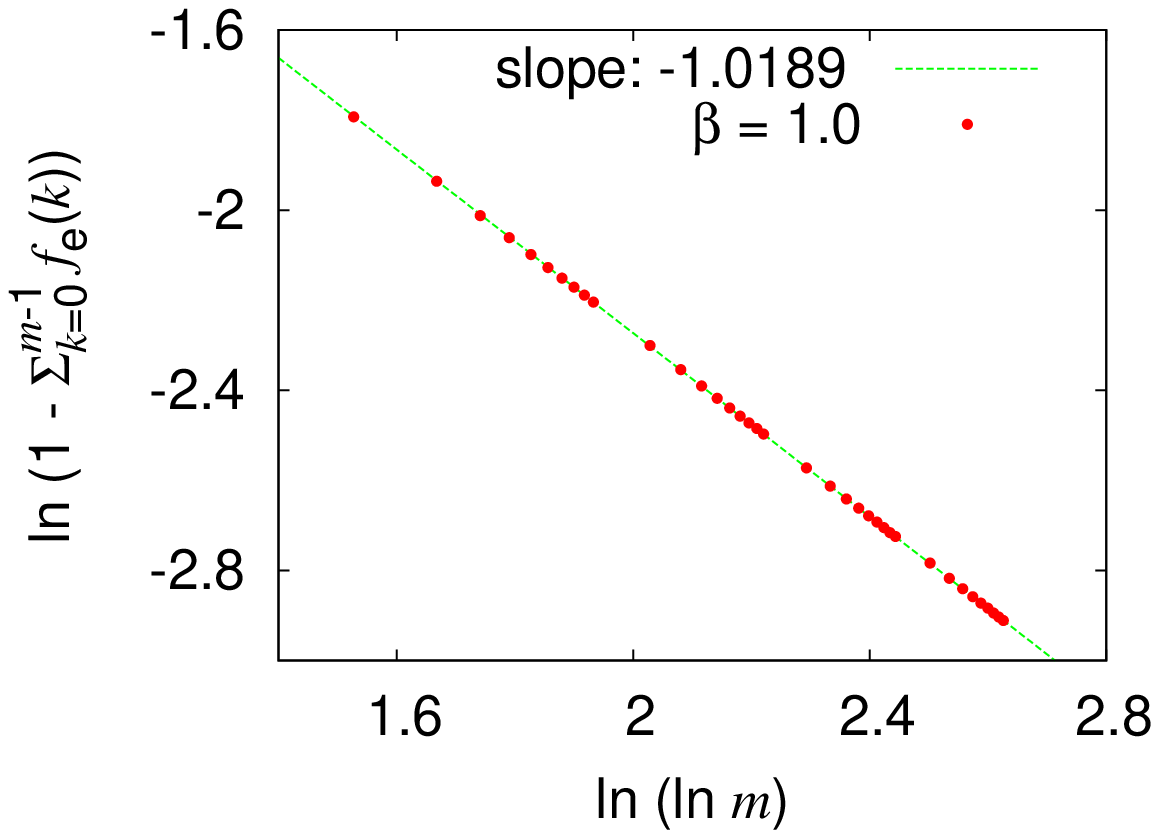}
		\label{fig4a}
	}
	\quad
	\subfigure[]{
		\includegraphics[width=0.45\hsize]{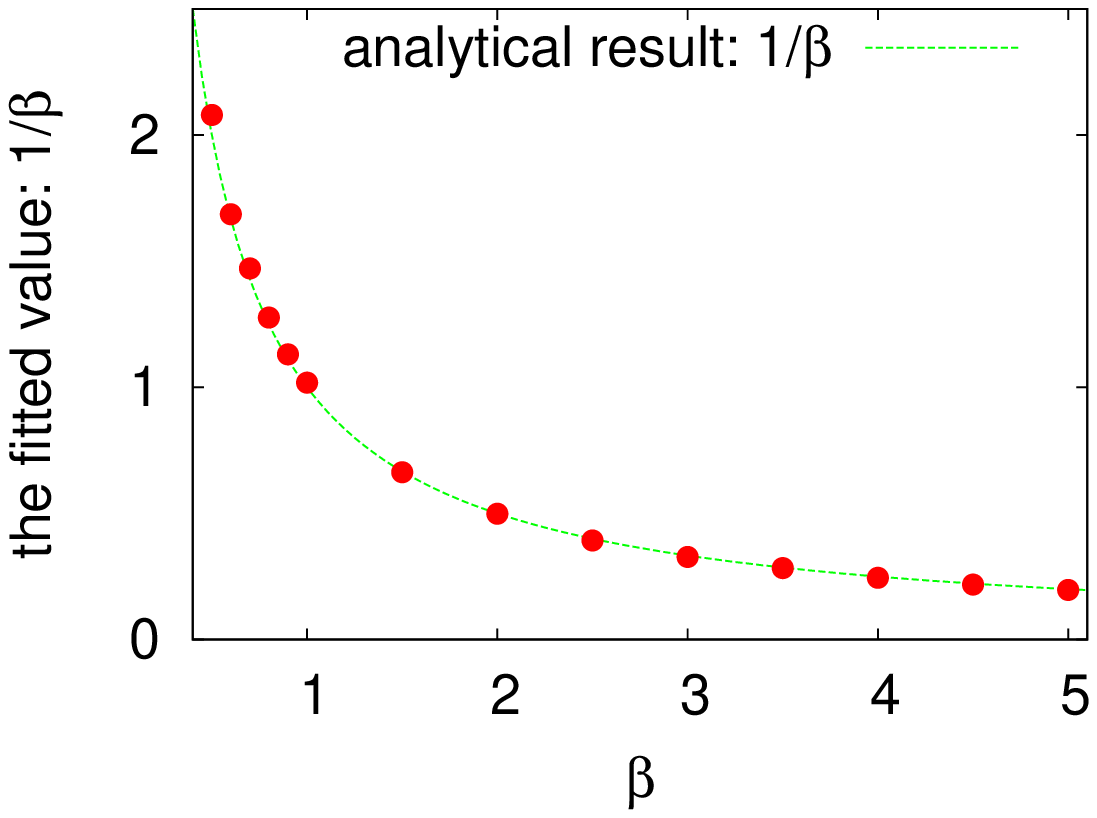}
		\label{fig4b}
	}
	\caption{(Color online) (a) Log-log graph of the survival probability $1 - \sum_{k=0}^{m-1} f_{\mathrm e}(k)$ versus $\ln m$.
	Points represent the numerical result for $\beta = 1.0$, and dashed line is obtained by least-squares fitting for large time-steps $(10^4 \le m \le 10^6)$.
	(b) Fitted -slope as a function of $\beta$.
	Dashed curve is the analytical result $1/\beta$.}
\end{figure}

From the definition of the first escape time distribution~(\ref{escape_measure}), the survival probability becomes
\begin{eqnarray}
	1 - \sum_{k=0}^{m-1} f_{\mathrm e}(k) & = & \int_0^{a_m} \, dx\nn\\
	& = & \mu(A_m) = O \left( (\ln m)^{-1/\beta}  \right) \quad \mbox{as} \quad m \to \infty. \label{survival}
\end{eqnarray}
Note that we can also obtain the same result by integrating eq.~(\ref{pdf_first_escape}).
Equation~(\ref{survival}) enables us to estimate the value $1 / \beta$ as
\begin{equation*}
	\ln \left(1 - \sum_{k=0}^{m-1} f_{\mathrm e}(k)\right) = -\frac{1}{\beta} \ln \,(\ln m) + \mbox{Const.}
\end{equation*}
for large $m$.
Figure~\ref{fig4a} shows a log-log graph of the survival probability versus $\ln m$ for $\beta = 1.0$, where $10^6$ initial points are given uniformly in the interval $(0,1)$.
The slope is obtained by least-squares fitting for large time-steps ($10^4 \le m \le 10^6$).
Figure~\ref{fig4b} shows the fitted -slope for $\beta$; it confirms the analytical result $1 / \beta$.

\begin{figure}
	\centering
	\includegraphics[width=\hsize]{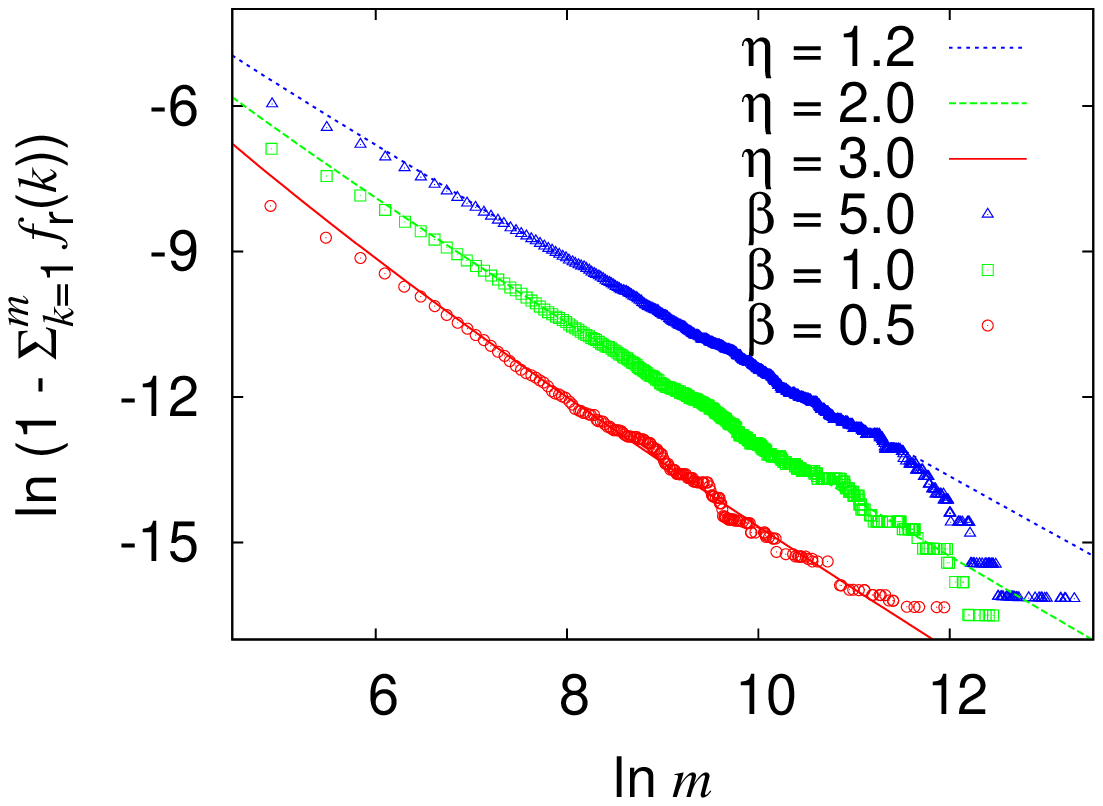}
	\caption{(Color online) Log-log graph of the probability $1 - \sum_{k=1}^m f_{\mathrm r}(k)$ versus $m$.
	Points represent the numerical results for $\beta = 0.5$ (circles), $1.0$ (squares), and $5.0$ (triangles).
	Each curve obeys eq.~(\ref{res_fitting_curve}) with $\eta = 1 + 1/\beta$ plus a constant.}
	\label{fig5}
\end{figure}

To estimate the residence time distribution, we calculated the probability $1 - \sum_{k=1}^m f_{\mathrm r}(k)$.
Equation~(\ref{pdf_residence}) implies that we can estimate the value $\eta$ as
\begin{equation}\label{res_fitting_curve}
	\ln \left( 1 - \sum_{k=1}^m f_{\mathrm r}(k) \right) = -\ln m - \eta \ln \, (\ln m) + \mbox{Const.}
\end{equation}
for large $m$.
Figure~\ref{fig5} shows a log-log graph of the probability versus $m$ for $\beta = 0.5,\, 1.0$, and $5.0$.
Each curve obeys eq.~(\ref{res_fitting_curve}) with $\eta = 1 + 1/\beta$ plus a constant.
For each $\beta$, good correspondence is observed around the region $7 < \ln m < 11$.

\subsubsection{Correlation function}
\begin{figure}
	\centering
	\subfigure[]{
		\includegraphics[width=0.45\hsize]{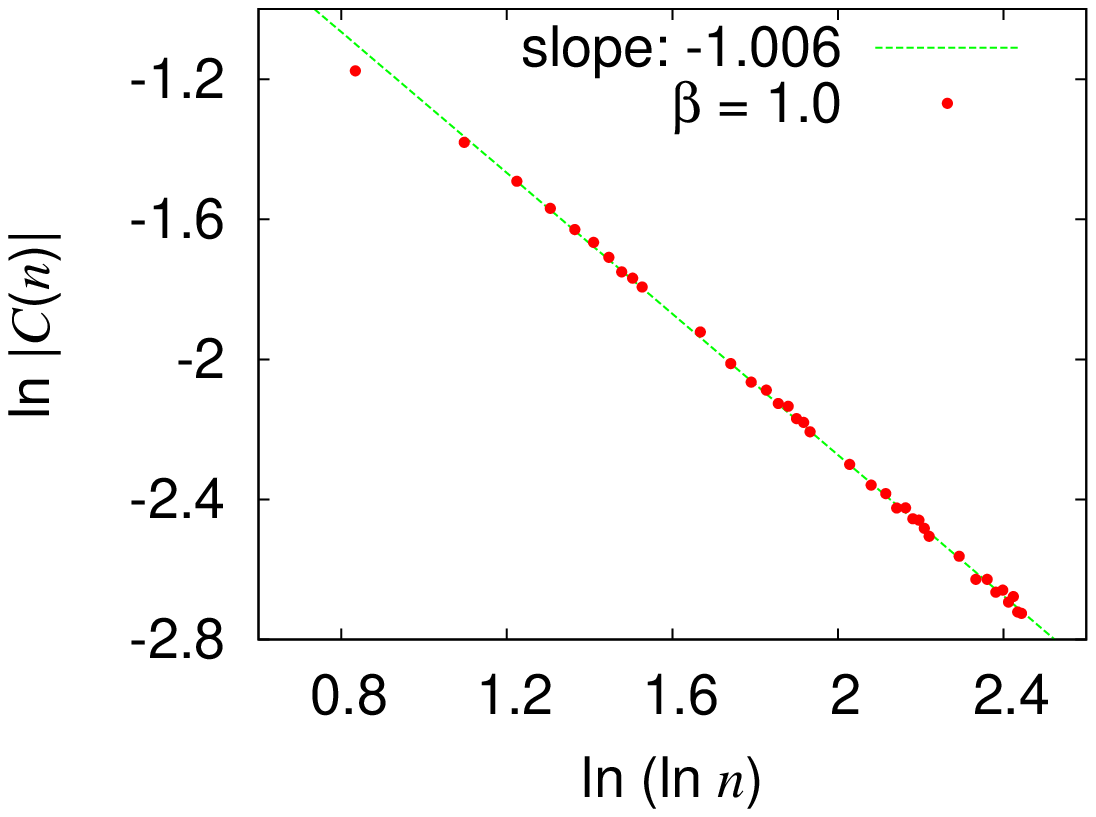}
		\label{fig6a}
	}
	\quad
	\subfigure[]{
		\includegraphics[width=0.45\hsize]{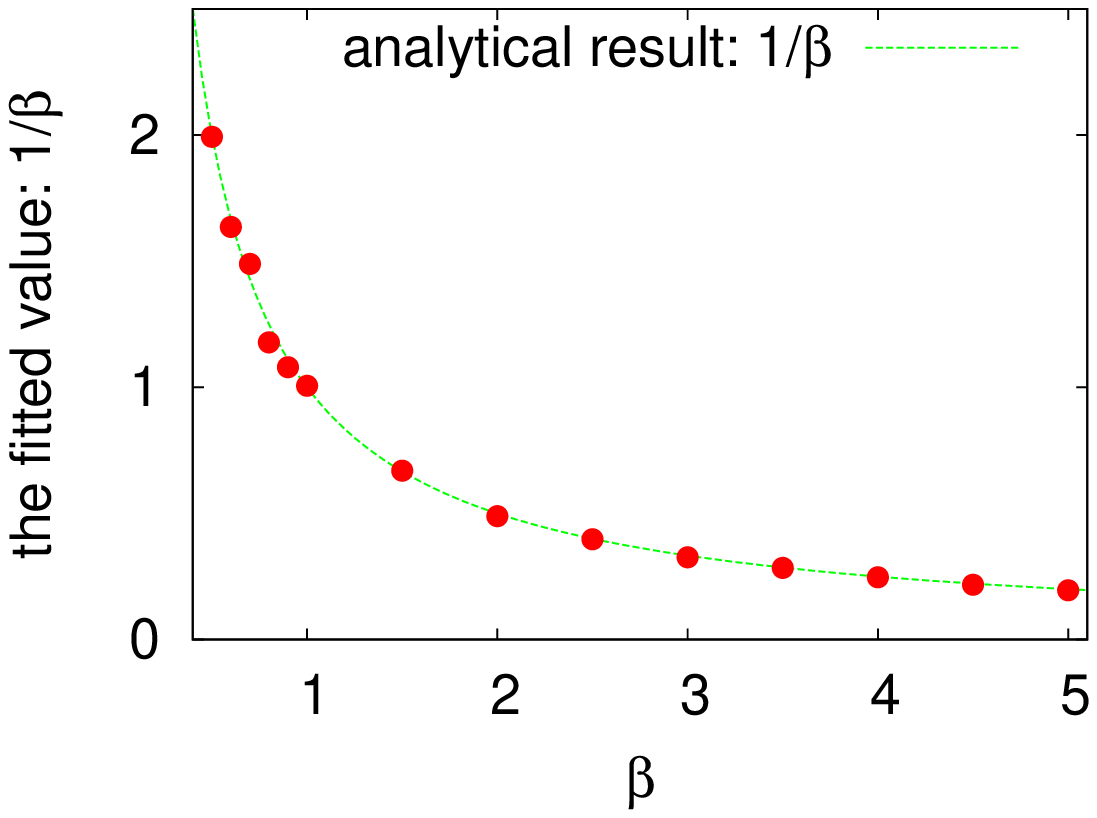}
		\label{fig6b}
	}
	\caption{(Color online) (a) Log-log graph of $|C(n)|$ versus $\ln n$.
	Points represent the numerical result for $\beta = 1.0$, and dashed line is obtained by least-squares fitting for large time-steps $(10^3 \le n \le 10^5)$.
	(b) Fitted -slope as a function of $\beta$.
	Dashed curve is the analytical result $1/\beta$.}
\end{figure}

In numerical simulations, we calculated the absolute value of the correlation function $|C(n)|$.
Equation~(\ref{correlation}) enables us to estimate the value $1/\beta$ as
\begin{equation*}
	\ln |C(n)| = - \frac{1}{\beta} \ln \, (\ln n) + \mbox{Const.}
\end{equation*}
for large $n$.
Figure~\ref{fig6a} shows a log-log graph of $|C(n)|$ versus $\ln n$ for $\beta = 1.0$, where $10^6$ initial points are given uniformly in the interval $(-1,1)$.
The slope is obtained by least-squares fitting for large time-steps ($10^3 \le n \le 10^5$).
Figure~\ref{fig6b} shows the fitted -slope for $\beta$; the analytical result $1 / \beta$ is verified.

\subsubsection{Power spectral density}
\begin{figure}
	\centering
	\includegraphics[width=\hsize]{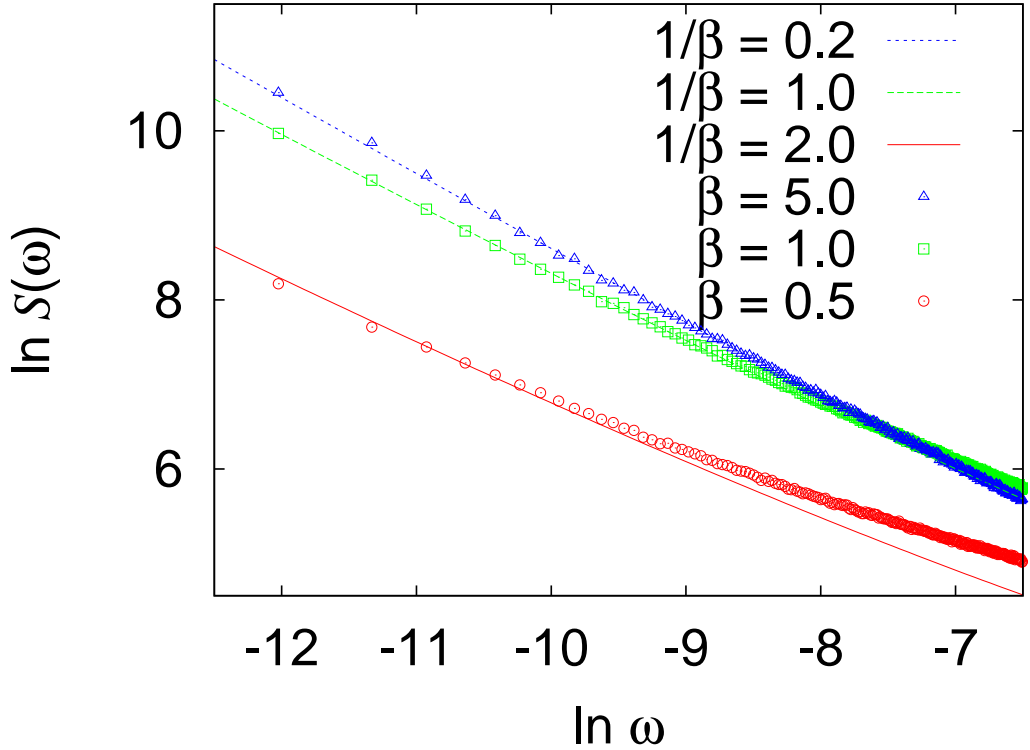}
	\caption{(Color online) Log-log graph of $S(\omega)$ versus $\omega$.
	Points represent the numerical results for $\beta = 0.5$ (circles), $1.0$ (squares), and $5.0$ (triangles).
	Each curve obeys eq.~(\ref{psd_fitting_curve}) with $1/\beta$ plus a constant.}
	\label{fig7}
\end{figure}

The power spectral density of coarse-grained orbits $\left\{ \sigma(U^n x) \right\}$ can be calculated as
\begin{equation*}
	S(\omega) = \frac{2 \pi \left\langle | \sigma^\ast(\omega_l) |^2 \right\rangle_\rho}{n},
\end{equation*}
where $\sigma^\ast(\omega_l)$ stands for the discrete Fourier transform defined by
\begin{equation*}
	\sigma^\ast(\omega_l) \equiv \sum_{k=0}^{n-1} \sigma (U^k x) \exp(- i \omega_l k) \qquad (\omega_l = 2 \pi l / n).
\end{equation*}
Our simulations were performed using the fast Fourier transform (FFT) method for $n = 2^{20}$ with $10^4$ initial points uniformly distributed in the interval $(-1,1)$.
Equation~(\ref{psd}) enables us to estimate the value $1/\beta$ as
\begin{equation}\label{psd_fitting_curve}
	\ln S(\omega) = - \ln \omega - \frac{1}{2\beta} \ln\left\{ (\ln \omega)^2 + \left(\frac{\pi}{2}\right)^2 \right\} +
	\ln \left[ \sin \left\{ \tan^{-1} \left(- \frac{\pi}{2 \ln \omega} \right) \Big/ \beta  \right\}  \right] + \mbox{Const.}
\end{equation}
for small $\omega$.
Figure~\ref{fig7} shows a log-log graph of $S(\omega)$ versus $\omega$ for $\beta = 0.5,\, 1.0$, and $5.0$.
Each curve obeys eq.~(\ref{psd_fitting_curve}) with $1/\beta$ plus a constant.
For $\beta = 1.0$ and $5.0$, we can observe good correspondence in the region $\ln \omega < -8$.
However, for $\beta=0.5$, the difference between the numerical result and the analytical curve is comparatively large.

\subsection{Distribution of partial sums}\label{ss4.3}
\begin{figure}
	\centering
	\includegraphics[width=\hsize]{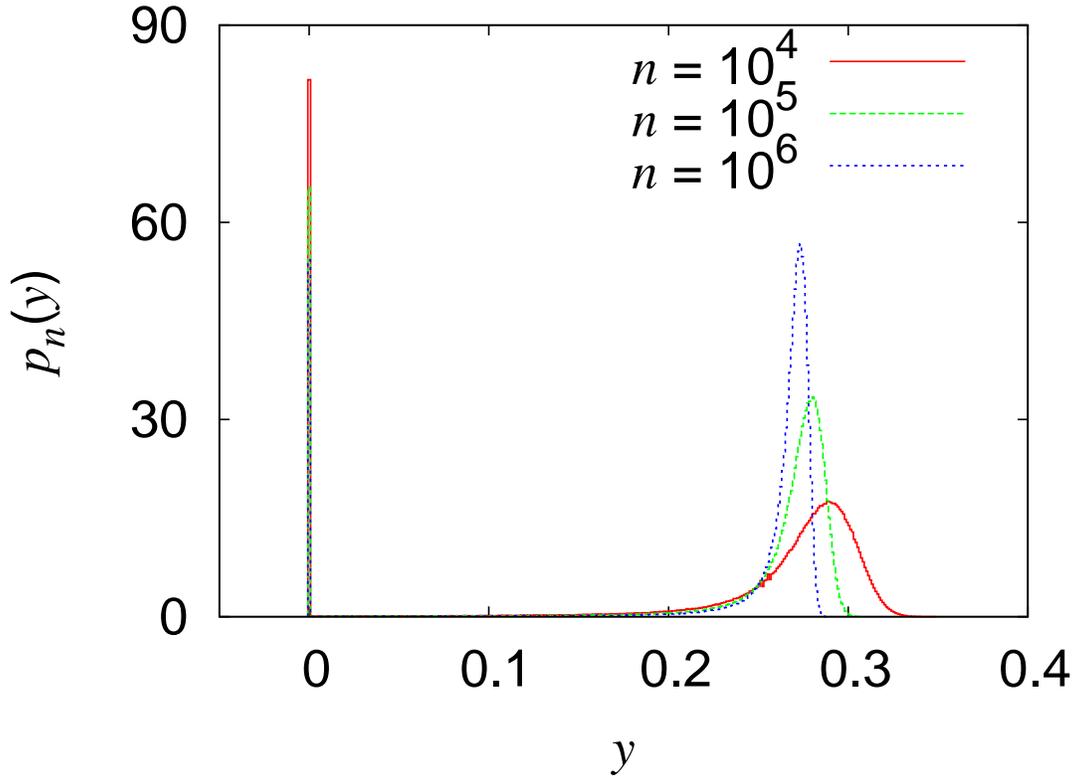}
	\caption{(Color online) Probability density $p_n(y)$ of $S_n(\phi) / n$ for $\beta = 1.0$ and $n=10^4$ (solid, red), $10^5$ (dashed, green), and $10^6$ (dotted, blue).}
	\label{fig8}
\end{figure}

Figure~\ref{fig8} shows the probability density $p_n(y)$ for $\beta = 1.0$ and $n = 10^4 \sim 10^6$.
In numerical simulations, $10^6$ initial points are provided in the interval $(0,1)$.
As mentioned in \S\ref{ss3.1}, the probability density is divided into two components.
One is observed at $y = 0$ and decays when $n$ increases.
The other asymmetrically converges to the phase average.
Our analytical result~(\ref{dis_partial_sum}) states that the first component decays as $(\ln n)^{-1/\beta}$, which is confirmed in Figs.~\ref{fig4a} and \ref{fig4b}, and the second component is the 1-stable probability density $\tilde{q}_n(1-y)$.
In what follows, we measure an exponent of $\tilde{q}_n$.

\begin{figure}
	\centering
	\includegraphics[width=\hsize]{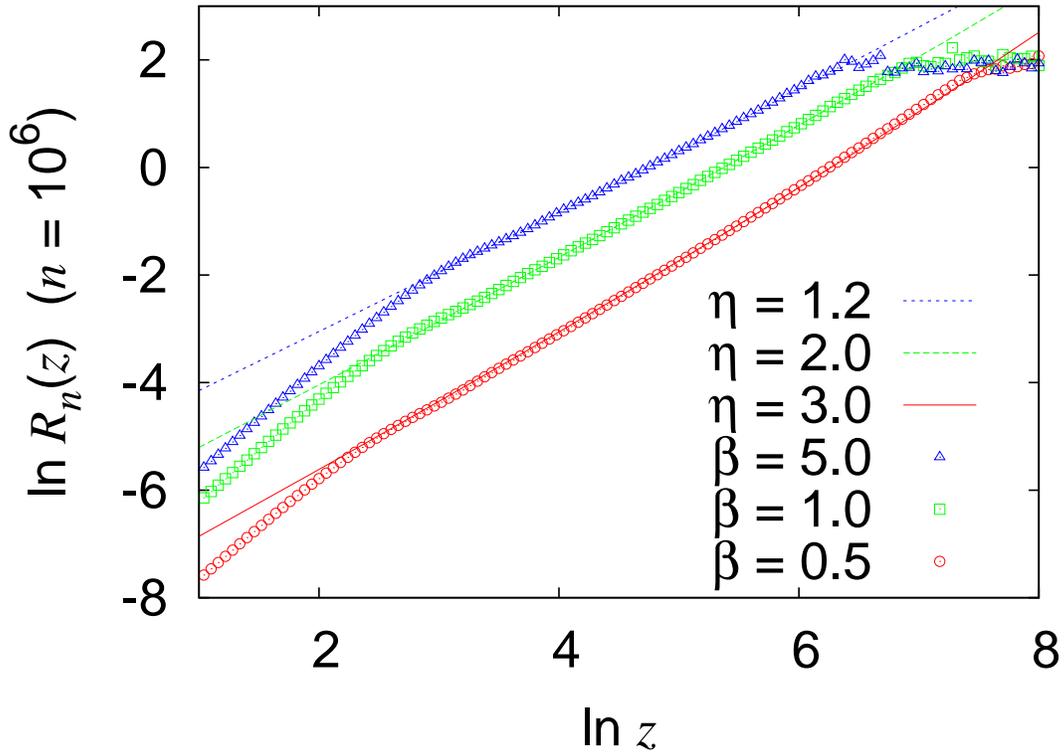}
	\caption{(Color online) Log-log graph of $R_n(z) \, (n=10^6)$ versus $z$.
	Points represent the numerical results for $\beta = 0.5$ (circles), $1.0$ (squares), and $5.0$ (triangles).
	Each curve obeys eq.~(\ref{Rnz_fitting_curve}) with $\eta = 1+1/\beta$ plus a constant.}
	\label{fig9}
\end{figure}

From eqs.~(\ref{qz}) and (\ref{C1}), the real part of $\ln \tilde{q}^\ast_n(z)$ becomes
\begin{equation*}
	R_n(z) \equiv - \mathrm{Re} \, \ln \tilde{q}^\ast_n(z) = O \left( z \left( - \ln \frac{z}{n} \right)^{-\eta} \right)
	\qquad (z > 0)
\end{equation*}
as $z / n \to 0$.
This enables us to estimate the exponent $\eta$ of $\tilde{q}_n$ as
\begin{equation}\label{Rnz_fitting_curve}
	\ln R_n(z) = \ln z - \eta \ln \, ( \ln n - \ln z) + \mbox{Const.}
\end{equation}
for small $z / n$.
Figure~\ref{fig9} shows a log-log graph of $R_n(z) \, (n=10^6)$ versus $z$ for $\beta = 0.5, \, 1.0$, and $5.0$.
Each curve obeys eq.~(\ref{Rnz_fitting_curve}) with $\eta=1+1/\beta$ plus a constant.
For each $\beta$, good correspondence is found around the region $3 < \ln z < 6$~\cite{footnote2}.

\begin{figure}
	\centering
	\subfigure[]{
		\includegraphics[width=0.45\hsize]{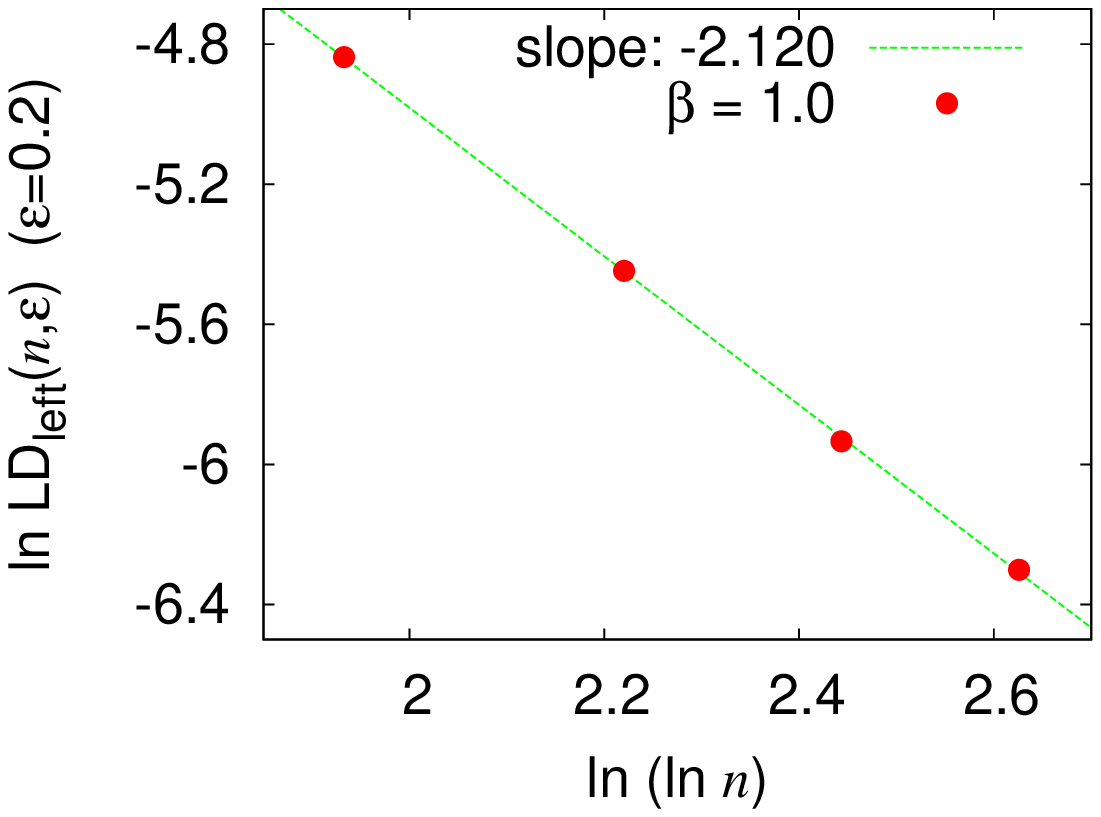}
		\label{fig10a}
	}
	\quad
	\subfigure[]{
		\includegraphics[width=0.45\hsize]{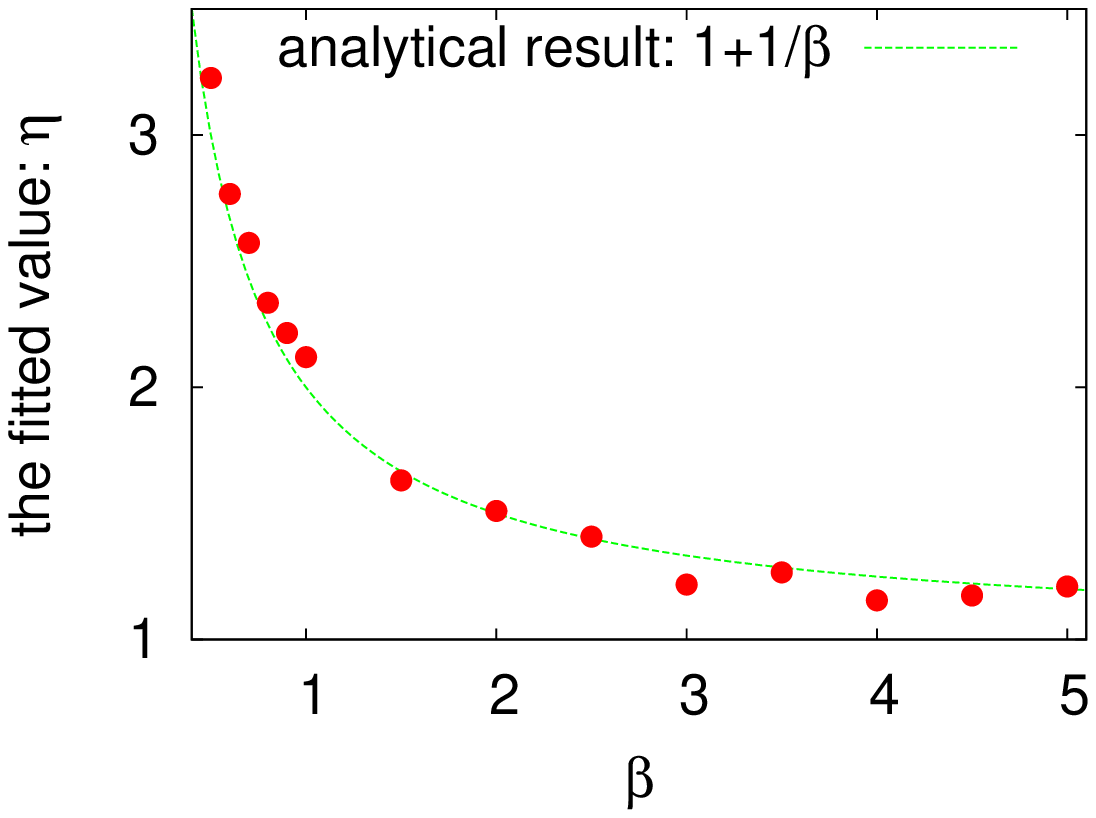}
		\label{fig10b}
	}
	\caption{(Color online) (a) Log-log graph of the large deviations versus $\ln n$.
	Points represent the numerical result for $\beta = 1.0$,
	and dashed line is obtained by least-squares fitting for large time-steps $(10^3 \le n \le 10^6)$.
	(b) Fitted -slope as a function of $\beta$.
	Dashed curve is the analytical result $\eta=1+1/\beta$.}
\end{figure}

In numerical simulations, we calculated the following large deviations,
\begin{equation*}
	\LD_\mathrm{left} (n,\varepsilon) \equiv \Pr \left\{ \frac{S_n(\phi)}{n} - \langle\phi\rangle_\rho < - \varepsilon \right\},
\end{equation*}
for a fixed constant $\varepsilon = 0.8 \langle\phi\rangle_\rho$ and uniformly distributed initial points $x \in A \setminus A_n$.
Equation~(\ref{large_deviations}) implies that the value $\eta$ of the large deviations can be estimated as
\begin{equation*}
	\ln \LD_\mathrm{left} (n,\varepsilon) = -\eta \ln (\ln n) + \mbox{Const.}
\end{equation*}
for large $n$.
Figure~\ref{fig10a} shows a log-log graph of the large deviations versus $\ln n$ for $\beta=1.0$ and $\varepsilon=0.2$.
The slope is obtained by least-squares fitting for large time-steps $(10^3 \le n \le 10^6)$.
Figure~\ref{fig10b} indicates that the numerical results tend to match the analytical result $\eta=1+1/\beta$.

\section{Summary and Discussion}\label{s5}
In this study, we introduced a class of intermittent maps having a uniform invariant measure and Nekhoroshev stability whose origin is the function $g$ defined by eq.~(\ref{def_g}).
Using renewal analysis, we showed that the correlation function decays according to the logarithmic inverse power law~(\ref{correlation}), and the power spectral density has the special form $1 / \omega$ with the logarithmic correction term~(\ref{eq:psd}).
The distribution of partial sums for the observable function $\phi$ consists of a delta peak and the 1-stable distribution~(\ref{dis_partial_sum});
the large deviations decay according to the logarithmic inverse power law~(\ref{large_deviations}).
The explicitly defined maps $T$, $U$, and the inverse function $g^{-1}$ enable us to calculate the dynamical properties numerically.
The numerical results clearly confirm our analytical results.

The observable function defined by eq.~(\ref{def_phi}) enables us to derive the 1-stable distribution by using renewal analysis.
In terms of ergodic theory, however, it is necessary to discuss not only the class of observable functions but also the limit distribution of sums for the functions.
This type of problem is an open problem~\cite{Sinai08}.
Gou\"ezel~\cite{Gouezel08} proved the limit theorem for the observable function $\phi(x)=x^{-\kappa}$ for the Bernoulli map $x \mapsto 2x \, \pmod{1}$.
Akimoto~\cite{Akimoto08} discussed a similar limit theorem for an infinite ergodic dynamical system.

To date, the logarithmic inverse power law of large deviations~(\ref{large_deviations}) is the slowest rate of decay:
an exponential decay is assumed in the large deviation principle~\cite{Ellis85}, and polynomial estimates have been obtained in the Pomeau-Manneville~\cite{Melbourne09,PollicottSharp09} and Pikovsky maps~\cite{ArtusoManchein09,Cristadoro10}.
In large deviation theory, not only polynomial estimates but also the logarithmic inverse power law are usually neglected, although $S_n(\phi) / n$ converges to the ensemble average.
One cannot calculate the entropy function defined by the Legendre transform of the logarithm of the moment-generating function (MGF), since the MGF of the 1-stable distribution cannot be defined.
Improving large deviation theory or statistical mechanics for our model is a problem that remains to be solved.

Finally, we note the relationship between our model and nearly-integrable Hamiltonian systems.
According to Niederman~\cite{Niederman04}, for the generic steep integrable Hamiltonian $H_0$, the constants $c_1$ and $c_2$ of Nekhoroshev stability~\cite{footnoteNekhoroshev} become $c_1 = c_2 \propto 1 / (2s)$.
Then, by comparing the log-Weibull laws in eqs.~(\ref{log_Weibull}) and (\ref{pdf_first_escape}), we can obtain
\begin{equation*}
	\beta = \frac{1}{s-1} \quad \mbox{and} \quad \eta = s.
\end{equation*}
In the study of chaotic Hamiltonian systems, a discussion of the relationship among the exponents and the degrees of freedom through the ergodic properties remains as future work.

\begin{acknowledgment}
This work is part of the outcome of research performed under a Waseda University Grant for Special Research Projects
(Project Number: 2009A-872).
\end{acknowledgment}

\appendix
\section{The Derivation of the Inverse Function $g^{-1}$ in Eq.~(\ref{def_g_inverse})}\label{AppA}
Let us consider the equation
\begin{equation*}
	y = g(x) = x^{1+\beta} \exp \left( 1 - x^{-\beta} \right),
\end{equation*}
and take the logarithm of both sides:
\begin{equation}\label{A1}
	\ln y = (1+\beta) \ln x + 1 - x^{-\beta}.
\end{equation}
For the new variables $t = x^{-\beta}$ and $Y = 1 - \ln y$, eq.~(\ref{A1}) becomes
\begin{equation}\label{A2}
	Y = \eta \ln t + t,
\end{equation}
where $\eta$ is defined by eq.~(\ref{def_eta}).
Equation~(\ref{A2}) can be transformed as follows:
\begin{equation}
	\frac{1}{\eta} \, e^{Y/\eta} = \frac{t}{\eta} \, e^{t/\eta}. \label{A3}
\end{equation}
Equation~(\ref{A3}) is in a special form that resembles the definition of the Lambert $W$ function, $z = W(z) \, e^{W(z)}$, so we can solve it as follows:
\begin{eqnarray}
	\frac{t}{\eta} & = & W \left( \eta^{-1} e^{Y/\eta} \right) \nn \\
	x^{-\beta} & = & \eta \, W \left( \eta^{-1} (e/y)^{1/\eta} \right) \nn \\
	x & = & \left\{ \eta \, W \left( \eta^{-1} (e/y)^{1/\eta} \right) \right\}^{-1/\beta}. \label{A4}
\end{eqnarray}
Equation~(\ref{A4}) shows that the inverse function of $g$ has the form
\begin{equation*}
	g^{-1}(t) = \left\{ \eta \, W \left( \eta^{-1} (e/t)^{1/\eta} \right) \right\}^{-1/\beta}.
\end{equation*}

\section{The Derivations of Eqs.~(\ref{pdf_first_escape}), (\ref{pdf_residence}), (\ref{correlation}), and (\ref{psd})}\label{AppB}
The function $G(x) =  - \int^x \frac{dy}{g(y)}$ and its inverse function are calculated as
\begin{equation}\label{G_Ginverse}
	G(x) = \frac{1}{e\beta} \, \exp \left( x^{-\beta} \right) \quad \mbox{and} \quad
	G^{-1}(m) = O \left( \left(\ln m\right)^{-1/\beta} \right)
\end{equation}
as $m \to \infty$, respectively.
If we use eqs.~(\ref{estimation_es_dis}) and (\ref{estimation_res_dis}), the probability density functions of the first escape and residence times become
\begin{eqnarray*}
	f_{\mathrm e}(m) & = & O \left( \left( (\ln m)^{-1/\beta} \right)^{1+\beta} \cdot
	\exp \left\{ - \left( (\ln m)^{-1/\beta} \right)^{-\beta} \right\} \right)\\
	& = & O \left( m^{-1} \, (\ln m)^{-\eta}  \right),\\
	f_{\mathrm r}(m) & = & O \left( \left\{ 1 + \eta \, \left( (\ln m)^{-1/\beta} \right)^{\beta}  \right\}
	\cdot \exp \left\{ - \left( (\ln m)^{-1/\beta} \right)^{-\beta} \right\} \cdot
	m^{-1} \, (\ln m)^{-\eta}  \right)\\
	& = & O \left( \left\{ 1 + \eta \, (\ln m)^{-1} \right\} \, m^{-2} \, (\ln m)^{-\eta} \right),
\end{eqnarray*}
as $m \to \infty$.
The exponent $\eta$ is defined by eq.~(\ref{def_eta}).
Relationship (\ref{relationship_fe_fr}) is obviously satisfied.

In calculating the Laplace transform $f_{\mathrm r}^\ast(z)$, for convenience, let us assume
\begin{equation*}
	f_{\mathrm r}(m) = e \left[ 1 + \eta \left\{ (\ln (m+e) \right\}^{-1} \right] \, (m+e)^{-2} \,
	\left\{ \ln (m+e) \right\}^{-\eta},
\end{equation*}
for $m \ge 0$.
Moreover, functions that satisfy $F_{\mathrm r}'(m) = f_{\mathrm r}(m)$ and $\tilde{F}_{\mathrm r}'(m) = F_{\mathrm r}(m)$ are defined as
\begin{equation*}
	F_{\mathrm r}(m) = - e (m+e)^{-1} \left\{ \ln (m+e) \right\}^{-\eta} \quad \mbox{and} \quad
	\tilde{F}_{\mathrm r}(m) = e \beta \left\{ \ln (m+e) \right\}^{-1/\beta}.
\end{equation*}
Then, the first moment $\langle m \rangle_{\mathrm r} = \int_0^\infty m f_{\mathrm r}(m) \, dm$ exists and becomes equal to $e \beta$;
for $\beta > 0$, however, the second moment $\langle m^2 \rangle_{\mathrm r}$ is infinite.
Using the partial integral, we can derive the Laplace transform as
\begin{eqnarray}
	f_{\mathrm r}^\ast (z) & = & \int_0^\infty e^{-zm} f_{\mathrm r}(m) \, dm \nn\\
	& = & \left[ e^{-zm} F_{\mathrm r}(m) \right]_0^\infty + z \, \left[ e^{-zm} \tilde{F}_{\mathrm r}(m) \right]_0^\infty + z^2 \,
	\int_0^\infty e^{-zm} \tilde{F}_{\mathrm r}(m) \, dm \nn\\
	& = & 1 - \langle m \rangle_{\mathrm r} \, z + 
	\langle m \rangle_{\mathrm r} \, z^2 \, \int_0^\infty e^{-zm} \left\{ \ln (m+e) \right\}^{-1/\beta} \, dm.
	\label{B1}
\end{eqnarray}
By applying the Tauberian theorem to eq.~(\ref{B1}), the Laplace transform can be estimated as
\begin{equation}
	f_{\mathrm r}^\ast (z)  =  1 - \langle m \rangle_{\mathrm r} \, z +
	\langle m \rangle_{\mathrm r} \, z \, (- \ln z )^{-1/\beta} + o \left( z \, (- \ln z )^{-1/\beta} \right) \label{B2}
\end{equation}
as $z \to 0$.

From eqs.~(\ref{Laplace_correlation}) and (\ref{B2}), the Laplace transform of the correlation function for the equilibrium renewal process can be calculated as
\begin{equation*}
	C^\ast (z) \sim z^{-1} \, (- \ln z)^{-1/\beta}
\end{equation*}
as $z \to 0$.
If the Tauberian theorem is applied again, the inverse transformation leads to
\begin{equation*}
	C(n) \sim (\ln n)^{-1/\beta}
\end{equation*}
as $n \to \infty$.

As mentioned in \S\ref{ss2.3}, the Wiener-Khinchin theorem provides that the power spectral density is defined by the real part of $C^\ast(i\omega)$.
Let us assume that a solution of the equation $e^{iz} = i$ is the principal value $z = \pi / 2$.
Then, $C^\ast \left( e^{i \frac{\pi}{2}} \omega \right)$ are calculated as
\begin{equation*}
	C^\ast \left( e^{i \frac{\pi}{2}} \omega \right) \sim e^{i\left( \frac{\theta(\omega)}{\beta} - \frac{\pi}{2} \right)} \,
	\omega^{-1} \, \left\{ r(\omega) \right\}^{-1/\beta}
\end{equation*}
as $\omega \to 0$, where the functions $r(\omega)$ and $\theta(\omega)$ are defined by eq.~(\ref{r_theta_omega}).
This asymptotic relation leads to
\begin{equation*}
	S(\omega) \sim \omega^{-1} \, \left\{ r(\omega) \right\}^{-1/\beta} \, \sin \left( \theta(\omega) / \beta \right)
\end{equation*}
as $\omega \to 0$.

\section{The 1-Stable Distribution with the Slowly Varying Function $(\ln x)^{-\eta}$}\label{AppC}
According to the theory of stable laws~\cite{Feller71}, the residence time distribution~(\ref{pdf_residence}) belongs to the domain of attraction of the so-called {\em 1-stable law} or {\em distribution}~\cite{AaronsonDenker98}, since the distribution satisfies
\begin{eqnarray*}
	\Pr \{ \bmm_j > x \} & = & \int_x^\infty f_{\mathrm r}(m) \, dm\\
	& = & \left( c_3 + o(1) \right) \, x^{-1} \, (\ln x)^{-\eta} \quad \mbox{as} \quad x \to \infty,
\end{eqnarray*}
where $c_3$ is a normalized constant, and it is obvious that the function $(\ln x)^{-\eta}$ is slowly varying at $\infty$.
The real and imaginary parts of the logarithm of its characteristic function $\psi(z) \equiv \int e^{izm} f_{\mathrm r}(m) \, dm$ can be estimated as follows~\cite{AaronsonDenker98}:
\begin{eqnarray}
	\mathrm{Re} \, \ln \psi(z) & = & - \frac{\pi}{2} \, c_3 |z|  (-\ln |z|)^{-\eta} + o \left( |z|  (-\ln |z|)^{-\eta} \right),\label{C1}\\
	\mathrm{Im} \, \ln \psi(z) & = & \gamma z + c_3 c_4  z  (-\ln |z|)^{-\eta} + z  K\left( |z|^{-1} \right)
	+ o \left( |z|  (-\ln |z|)^{-\eta} \right),\nn
\end{eqnarray}
as $z \to 0$, where the function $K$ and the constants $c_4, \, \gamma$ are defined by
\begin{eqnarray*}
	K(y) & = & \int_0^y \frac{\left( c_3 + o(1) \right) \, t \, (\ln t)^{-\eta}}{1+t^2} \, dt ,\\
	c_4 & = & \int_0^\infty \left( \cos t - \frac{1}{1+t^2} \right) \, \frac{dt}{t},\\
	\gamma & = & \int_0^\infty \left( \frac{t}{1+t^2} + \mathrm{sgn} (t) \int_0^{|t|} \frac{2u^2}{(1+u^2)^2} \, du \right)
	f_{\mathrm r}(t) \, dt.
\end{eqnarray*}



\begin{thebibliography}{99}
\bibitem{footnoteNekhoroshev}
The so-called Nekhoroshev stability in nearly-Hamiltonian systems states that, for every initial point, the motion is confined around $I(0)$ for an exponentially long time,
\begin{equation*}
	\left| I(t) -I(0) \right| < \varepsilon^{c_1} \quad \mbox{for} \quad |t| < \varepsilon^{-1} \exp \left( \varepsilon^{-c_2} \right),
\end{equation*}
where $\varepsilon$ stands for a perturbation parameter, and $c_1$ and $c_2$ are positive constants determined by the steepness condition of the integrable Hamiltonian.
\bibitem{footnote1}
The function $g(t) = t^B \, (B>1)$ also satisfies three conditions.
In this case, the map $T$ is approximately equal to the Pikovsky map with a parameter $z$~\cite{Pikovsky91,Cristadoro10} and the Miyaguchi map with a parameter $\beta$~\cite{MiyaguchiAizawa07}.
These parameters correspond to $z = B$ and $\beta = B/(B-1)$, respectively.
\bibitem{footnote2}
In Fig.~\ref{fig9}, numerical results are out of theoretical curves for $\ln z > \ln z_\mathrm{l}$ and $\ln z < \ln z_\mathrm{s}$ regions.
Due to the factor $e^{izy}$ in calculating the characteristic function,
the minimum and maximum sizes of the numerically obtained probability density $\tilde{q}_n$ correspond to the half wavelengths $\pi / z_\mathrm{l}$ and $\pi / z_\mathrm{s}$, respectively.
\end{thebibliography}
\end{document}